\documentstyle[12pt]{article}
\begin{document}
\baselineskip 6mm
\begin{center}
\Large{{\bf
Quantization of
Gravitational Waves and
the Squeezing}}
\end{center}
\begin{center}
I. Lovas       \\
 Department of Theoretical Physics,
 \\  Kossuth  Lajos University,     \\
H-4010 Debrecen P.O.Box 5, Hungary\\
\end{center}

{\bf Abstract}
 The question whether gravitational waves are
quantised or not can in principle  be answered by the help of
correlation measurements.
If the gravitational waves are quantised
and they are
generated by the change of the background metrics then they can be in
squeezed state.
 In a sqeezed state there is a  correlation
between the phase of the wave and  the quantum fluctuations. It is
recommended to analyse the data to be obtained  by the
gravitational detectors  from the point of view of such
correlations.
An explicit formula is derived for the squeezing parameter of the
quantised gravitational waves. The head on collision of two
identical black holes is analysed as a possible source of
squeezed gravitational waves.
 
\section{ Introduction}
\indent
 
  A great deal of  efforts are devoted  for the unification of
the quantum theory and
the theory of general relativity.
In addition to this formidable theoretical task, there is a
similar
experimental problem: how to observe the presence of  quantum
effects in gravitational phenomena. This is the question, that we
want to investigate  in this letter. It is an old
belief
       that
 the
 periodic distortion of the curvature  can propagate
in  space-time as a gravitational wave \cite{Zac73} . The observations
 conducted
by Hulse and Taylor \cite{Tay75}
provided a convincing proof for this beliefe.
 The periastron shift of the double neutron star
 PRS 1916+13
is
in  excellent agreement with the calculations based on the
Einstein-equations \cite{Dam86}  proving         that the double star  is
loosing   its  energy continuously
by the emission  of gravitational waves.
 
At the moment there are a number of laboratories on all over
  the world where detectors are
planed or built for the observation of  gravitational waves.
 In addition to  Weber-type resonators \cite {Web61},
 large Michelson-interferometers, as the VIRGO in Europe and the LIGO
in USA are being built.
It is assumed that we shall have already at the beginning of the next
decade observational evidences
for the existence of the gravitational waves.
 
     It is also an old belife that the gravitational waves are
quantized. The quantised gravitational wave, the graviton, is a
massless, transversely polarized
particle carying energy of  $\hbar \omega$ and spin 2 $\hbar$.
This
belife  is based only on analogies and not on a well established
theory contrary to the gravitational waves which were predicted
   theoretically 80 years ago \cite{Ein18}.
The situation is  even worst as far as the observational
 possibilities
are concerned. The frequency of the gravitational waves must be very
low $(10 Hz-10 kHz)$
if they are  emitted by  heavy stellar objects, consequently the energy
of the gravitons is extremely low.
To observe quantum effects of the gravitons, in the original sense,
i.e.
the analogue of the photoeffect, is completely hopeless, therefore
we must focus our
attention
to many graviton systems and look for phenomena which are
sensitive to the bosonic character of the gravitons.
 
 There was a famous,
discovery in the radioastronomy made by Hanbury Brown and Twiss
 \cite{Han56} in the middle of the fifties.
  Since the
many photon state must be symmetric, the correlation
function
of two photons observed in coincidence by two radiotelescope
exhibits an enhancement at small values of
$q=|{\bf (k_1-k_2)}|$,       where the momenta of the photons
are denoted
 by ${\bf k_1}$ and ${\bf k_2}$.
This is a consequence of the bosonic nature of the photons,
therefore it can be considered as a proof of the quantised
behaviour of the elactromagnetic field.
      The correlation functions of the electromagnetic radiation
were
studied by Glauber  \cite{Gla63} in the
 framework of  quantumelectrodynamics.
In order to clarify the problems of the
correlations he used  the notion
of the coherent state $|\alpha_k>$     defined by
 
\begin{eqnarray}
 |\alpha_k>= exp(-|\alpha_k|^2/2) {\large \Sigma_{n_k}}
\frac{\alpha_k^{n_k}} {\sqrt{n_k !}} |n_k>,
\end{eqnarray}
where  the wave vector and  the
polarization vector is denoted by the compact index $k$.
Glauber has proved that the correlation function
is different from 1,
if the light is a statistical
mixture or a superposition of coherent states, however, it
reduces to 1 if the
light is completely coherent.
 
    It is important to point out, that
the detectors of the gravitational waves,  are sensitive
only
to completely coherent states, thus we are unable to exploit
the analogue of the Hanbury Brown and Twiss effect to prove the
quantised character of the gravitational waves.
 
       We must look for
an other possibility. In the following we are going to investigate
the possibility of the "squeezing", which is a "genuin" quantum
phenon in the realm of the electromagnetic waves.\\
 
\section{ Squeezed Gravitational Waves}
\indent
 
 A squeezed state of light can be generated from the vacuum state
by the action of the operator S defined as follows
\begin{eqnarray}
S= exp \left[z_{k}^* a_{+k} a_{-k} - z_{k}
  a_{+k}^+ a_{-k}^+ \right].
\end{eqnarray}
where
$z_k$ is a complex parameter. The most
remarkable property
of the squeezed state is the correlation between the quantum
fluctuations and the phase of the wave.\\
 
      The gravitational waves may be in squeezed state.
The proof
of this statemant
\cite{Gri90} can be summarised as follows.
The metric tensor is decomposed
into a slowly varying  backgound and a rapidly fluctuating
infinitesimal part:
\begin{eqnarray}
 g_{m n} (x)= g^{0}_{m n} (x) +h_{m n} (x)
\end{eqnarray}
In this  case the Einstein-equation can be reduced to a wave equation:
\begin{eqnarray}
D_l D^l h_{m n} (x)=0,
\end{eqnarray}
where the covariant derivative, denoted by $D_l$, depends
 on the background metrics
 $g^{0}_{m n} (x).$
 The
solution $h_{m n}(x)$ must satisfy the gauge fixing condition
given by
\begin{eqnarray}
\frac12  D_m h_l^l (x)= D_l h_m^l(x).
\end{eqnarray}
The wave equation and the gauge fixing condition are linear,
 consequently the field $h_{m n} (x)$ can be expanded at a
given time $t$ in terms of the complete, orthonormal set of functions
 $v_{m n} (k,\vec{x})$
 satisfying the wave equation and the gauge fixing condition given
above:
\begin{eqnarray}
h_{m n} (x) = \mathop{\sum_{k}} [ a_k
v_{m n}(k,\vec{x}) e^{i\omega_{k} t} +
a^+_k v^*_{m
  n}(k,\vec{x}) e^{-i\omega_k t}],
\end{eqnarray}
In the canonical quantisation approach
 $a_k$ and $a^+_k$
 must satisfy
the commutation relations given by
\begin{eqnarray}
\left[a_k , a_{k'}\right]=0,
\left[a^+_k ,
  a^+_{k'}\right]=0, \left[a^+_k,
  a_{k'}\right]=\delta_{k k'}.
\end{eqnarray}
The vacuum state $|\Psi(0)>$ at time $t=0$ is defined by the help
of the graviton annihilation operator
$a_k$:
\begin{eqnarray}
a_k|\Psi(0)> = 0.
\end{eqnarray}
Acting repeatedly on the vacuum state by the graviton creation
operator $a^+_k$   $n$-particle
states are generated and a Fock-space can be constructed.
At a later time $\delta t$ the vacuum state $|\Psi(0)>$ defined at
$t=0$ is
not "empty" in general, since the change of the background metrics
$g^0_{m n} (x)$ during the time $\delta t$
mixes the positive and
negative frequency components,
or in other words mixes the creation and annihilation operators.
The new annihilation operator $b_{k}$
 can be obtained by the following
Bogoljubov-Valatin transformation
\cite{Val61}:
\begin{eqnarray}
b_{k}= \alpha_{k} a_{k}+ \beta_{k}^*
a_{-k}^+ .
\end{eqnarray}
The expectation value of the number operator $N_k$ at time $
\delta t$ is given by
\begin{eqnarray}
<\Psi(0)| b_{k}^ + b_{k}|\Psi(0)> = |\beta_{k}|^2 \neq 0.
\end{eqnarray}
Now it is possible to show that the gravitons produced by the
change of the background metrics are in squeezed state. First of
all we define a unitary operator  which transform
$a_{k}$ into $b_{k}$:
\begin{eqnarray}
b_{k} = U a_{k} U^+.
\end{eqnarray}
The time development operator $U$ can be written in the
following form:
\begin{eqnarray}
U = R S,
\end{eqnarray}
where the operator of the "rotation" R and that of the
"squeezing" are defined by
\begin{eqnarray}
R = exp \left[-i \vartheta_{k} \left( a_{+k}^+ a_{+k} + a_{-k}^+
  a_{-k}\right)\right]
\end{eqnarray}
and
\begin{eqnarray}
S= exp \left[z_{k}^* a_{+k} a_{-k} - z_{k}
  a_{+k}^+ a_{-k}^+ \right],
\end{eqnarray}
respectively.
Here the parameters $
 \vartheta_{k}$,
 $r_{k}$ and $ \phi_{k}$ defined by
 $z_{k}= r_{k} e^{2 i \phi_{k}}$ are real
numbers. It is possible to show that the paramaters of the
Bogoljubov-Valatin transformation $\alpha_{k}$ and
$\beta_{k}$
can be expressed in terms of the parameters of the operator $U$
by the help of the
following relations:
\begin{eqnarray}
\alpha_{k}&=& exp \left(i \vartheta_{k}\right) cosh \left(r_{k}\right)\\
\beta_{k}&=& exp \left[ i(\vartheta_{k} - 2 \phi_{k})\right] sin
h\; \left(r_{k}\right).
\end{eqnarray}
Here the important point is the fact that $S$ is a
sqeezing operator. The time development of the state of the
graviton system is generated by a squeezing operator.
 
In the rest of this paper we will show  that
  the time development operator can be
calculated for any finite time $t$ and
an explicit expression can be
derived for the squeezing parameter $z_{k}$.
    The total four momentum $P^{i}$ of the physical system can be
expressed as
\begin{eqnarray}
P^{i} =\frac{1}{c} \int  (-g) ( T^{ij} (x) + t^{ij}(x)) d S_{j}
\end{eqnarray}
where $g$ is the determinant of $g_{ij}$, the
energy--momentum tensor of matter is denoted by $T^{ij} (x)$
and the energy--momentum pseudotensor $t^{ij} (x)$ of the curved
space is defined as:
\begin{eqnarray}
t^{ij}(x) = \frac{c^4}{16 \pi G \sqrt{(- g)}} \partial_{l}
\partial_{m}
{ \sqrt{(-g)} (g^{ij} (x) g^{lm} (x) - g^{il}(x)
g^{jm}(x)) }.
\end{eqnarray}
In the absence of matter the energy-momentum tensor  $T^{ij}(x)$
vanishes:
\begin{eqnarray}
  T^{ij}(x) = 0
\end{eqnarray}
and the energy of the system can be written in the following form:
\begin{eqnarray}
 H[g_{ij}] = \int {\cal H} (g_{ij}(x))
 d^3x
\end{eqnarray}
where
\begin{eqnarray}
{\cal H}(g_{ij}(x)) =\frac{1}{c}  (-g(x))
t^{00}(x).
\end{eqnarray}
The energy of the curved space expanded around $g^0_{ij}(x)$ can be
expressed as a power series of $h_{ij}(x)$:
\begin{eqnarray}
 H[g_{ij}] = \int ({\cal H}^0(x)+ (\frac{\delta {\cal H}}
{\delta g_{ij}})^0
h_{ij}(x)\\+
 \frac{1}{2} (\frac{\delta^2 {\cal H}}{\delta (g_{ij})^2})^0
(h_{ij}(x))^2+ ... )
d^3x
\end{eqnarray}
where the superscript $0$ indicates that the expansion
coefficients are calculated
from the background metrics $g^0_{ij}$. It has to be noted that
the terms
linear or bilinear in $h_{ij}$ do not give contribution to the integral.
 The interaction
energy between the
background and the gravitational waves can be written as follows:
\begin{eqnarray}
H'(t) = H[g_{ij}]-H[g_{ij}^0]=
  \frac{1}{2} \int (\frac{\delta^2 {\cal H}}{\delta (g_{ij})^2})^0
(h_{ij}(x))^2 d^3x.
\end{eqnarray}
Introducing the expansion for $h_{ij}(x)$ defined by Equ.(6)
we obtain the
 following expression:
\begin{eqnarray}
 H'(t) =\frac{1}{2}\int
(\frac{\delta^2 {\cal H}}{\delta (g_{ij})^2})^0
(\mathop{\sum_{k}} (a_k a_{-k} v_{ij}(k,\vec{x}) v_{ij}(-k,\vec{x})
  exp(2i \omega_k t)\\ +
       (a_k^{+} a_{-k}^{+} v^*_{ij}(k,\vec{x}) v^*_{ij}(-k,\vec{x})
 exp(-2i \omega_k t)\\ +
(a^+_k a_k+a^+_{-k} a_{-k}) v_{ij}(k, \vec{x})
v^*_{ij}(k,\vec{x})) d^3x.
\end{eqnarray}
 
By the help of the interaction operator $H'(t)$, the state
vector $|\Psi(t)>$
of the gravitons at time $t$ can be expressed in the following
form:
 
\begin{eqnarray}
|\Psi(t)>=exp(
 \mathop{\sum_{k}}
  \left[z_{k}^* a_{+k} a_{-k} - z_{k}
  a_{+k}^+ a_{-k}^+ - i \vartheta_k
(a_{-k}^+a_{-k}+a_k^+a_k)\right]) |\Psi(0)>,
\end{eqnarray}
where
\begin{eqnarray}
z_k(t)&=& \frac{i}{2 \hbar}\int_0^t (\frac{\delta^2 {\cal
H}}{\delta
(g_{ij})^2})^0 v^*_{ij}(k, \vec{x}) v^*_{ij}(-k, \vec{x})
 exp(-2i \omega_k t')d^3x dt',\\
\vartheta_k(t)&=&  \frac{1}{2 \hbar}\int_0^t (\frac{\delta^2
{\cal H}}{\delta
(g_{ij})^2})^0 v_{ij}(k, \vec{x}) v^*_{ij}(k, \vec{x})
 d^3x dt'.
\end{eqnarray}
Using the squeezing parameter $z_k$ the correlation between the
phase of
the wave labelled by $k$ and the quantum noise can be calculated.

\section{ Collision of two Black Holes}
\indent
 
Here we investigate the most simple source of
gravitational waves, namely the
 head on collision of two identical, non rotating black
holes. In this case it is possible to solve numerically the
Einstein-equations
and furthermore it is possible to separate the background metrics $g^0_{ij}$.
Let us consider a curved space-time characterised by the line
element $ds$ defined as follows
\begin{eqnarray} d^2s&=&-dt^2+
dl^2,    \\
dl^2&=&\psi_M^4(\varrho, z)(d\varrho^2+dz^2+\varrho^2d\phi^2),
\end{eqnarray}
where the cylindrical coordinates are denoted by
 $\varrho, \phi$  and $z$. It was proved by
Misner \cite{Mis60}  that the Einstein-equations can be
satisfied if the function
$\psi_M(\varrho, z)$       is defined as
\begin{eqnarray}
\psi_M(\varrho, z)
=1+\sum^{\infty}_{n=1} \frac1{\sinh (n\mu)}
    \left[ \frac1{\sqrt{\varrho^2+(z+{\rm coth}n\mu)^2}} +
           \frac1{\sqrt{\varrho^2+(z-{\rm coth}n\mu)^2}} \right]\;
\end{eqnarray}
This corresponds to two, non rotating, equal mass M black holes,
siting at
\begin{eqnarray}
z=\pm {\rm coth} \mu\;.
\end{eqnarray}
Using this solution as an initial condition for the
collision of two identical black holes the time evolution of the
metrics can be followed by numerical solution of the
Einstein-equation \cite{Ann93}.
The two black holes in the initial state are  in rest,
then they begin to accelerate and finally the event horizonts
merge.
 Using the Regge-Wheeler perturbation theory
\cite{Reg57}
the time evolution of the gravitational waves can be
represented by the Zerilli-function
$ \psi^L_Z(r,t)$
\cite{Zer70},
which is the solution of the following equation:
\begin{eqnarray}
\frac{\partial^2 \psi^L_Z}{\partial t^2} -
\frac{\partial^2\psi^L_Z}{\partial r^2_*} +
V^L_Z(r) \psi^L_Z =0\;,
\end{eqnarray}
where
\begin{eqnarray}
r_*&=&r+M\;\ln \left(\frac r{2M} -1 \right)\;,\\
V^L_Z(r) &=& \left(1-\frac{2M}r \right)
 \frac{2\lambda^2(\lambda+1) r^3+6\lambda^2Mr^2+18\lambda M^2r+18M^3}
      {r^3(\lambda r+3M)^2}\\
 && \lambda= \frac12 (L-1)(L+2)\;.
\end{eqnarray}
By the help of the solution $\psi^L_Z(r,t)$
the numerically calculated metrics
$ g_{m n} (x)$ can be decomposed into $g^{0}_{m n} (x)$ and
$h_{m n} (x)$ and then the
squeezing parameters can in principle  be computed.
 
 According to the arguments given
above it seems to be worth while to
look for correlations between quantum noise and the phase
of the waves
in the data to be obtained by the
gravitational wave detectors.
\\
\[   \] \\
E-mail: lovas@heavy-ion.atomki.hu\\
\[   \]     \\

\vfill\eject


\begin{thebibliography}{99}
\itemsep -1mm
\parindent=8truemm \itemsep -1mm
\bibitem{Zac73} V.D.Zacharov, Gravitational waves in Einstein's
                theory, John Wiley and Sons (New York) 1973
\bibitem{Tay75} R.A.Hulse  and J.H.Taylor,   Astrophys. J. {\bf 195},
                L51  (1975)
\bibitem{Dam86} T.Damour and N.Deruelle, Ann. Inst. H.Poincar\'e
                (Phys. Th\'eor.) {\bf 44},  (1986) 263
\bibitem{Web61}  J. Weber, General Relativity and Gravitational Waves,
                 Interscience (New York) 1961
\bibitem{Ein18} A.Einstein, Preuss.Akad. der Wissenschaften
                {\bf 1}, (1918)  154
\bibitem{Han56}
            R.Hanbury Brown and R.Q.Twiss, Nature (London) {\bf 177},
             (1956)  27.\\
            R.Hanbury Brown and R.Q.Twiss, Nature (London) {\bf 178},
            (1956)  1046.
\bibitem{Gla63}   R. J. Glauber, Phys.Rev. Lett. {\bf 10},
                  (1963)  84.\\
                  R. J. Glauber, Phys.Rev. {\bf 131},  (1963) 2766.
\bibitem{Sto70}   D. Stoler,
                  Phys.Rev. {\bf D1},  (1970) 3217.\\
                  M. Rosenbluh and R.M. Shelby,
                  Phys.Rev. Lett. {\bf 66},  (1991) 153.
\bibitem{Gri90}
                  L. Grishchuk and Y.V. Sidorov,
                  Phys.Rev. {\bf D42}, (1990)  3414.
\bibitem{Val61}
                  N.N. Bogoljubov,
                  Nuovo Cimento {\bf 7}, (1958)  794.\\
                  J. G. Valatin,
                  Nuovo Cimento {\bf 7},  (1958) 843.
\bibitem{Mis60}   C.W. Misner, {\it Phys. Rev.} {\bf 118},
                   (1960) 1110.
\bibitem{Ann93}
                  P. Anninos, D. Habill, E.Seidel, L. Smarr and
                  Wai Mo Suen, {\it Phys. Rev. Letters} {\bf 71},
                   (1993) 2851.
\bibitem{Reg57}
                  T.Regge and J.A. Wheeler, {\it Phys. Rev.}
                  {\bf 108}, (1957)  1063.
\bibitem{Zer70}
                  F.J. Zerilli, {\it Phys. Rev. Letters} {\bf 24},
                   (1970) 737.
\end{thebibliography}
\end{document}